\documentclass[twocolumn,showpacs,preprintnumbers,amsmath,amssymb,prl]{revtex4}

\usepackage{dcolumn}
\usepackage{latexsym}
\usepackage{graphics}
\usepackage{color}
\usepackage{epsfig}
\usepackage{bm}

\begin{document}

\title{\boldmath Evidence for an Excited Hyperon State in $pp \rightarrow p K^+ Y^{0*}$}

\author{
  I.~Zychor,$^1$
  V.~Koptev,$^2$
  M.~B\"uscher,$^3$
  A.~Dzyuba,$^2$
  I. Keshelashvili,$^{3,4}$
  V.~Kleber,$^5$*
  H.R.~Koch,$^3$
  S.~Krewald,$^3$
  Y.~Maeda,$^3$
  S.~Mikirtichyants,$^2$
  M.~Nekipelov,$^{2,3}$
  H.~Str\"oher,$^3$
  C.~Wilkin,$^6$
}

\affiliation{$^1$The Andrzej So{\l}tan Institute for Nuclear Studies, 05400 \'Swierk, Poland}
\affiliation{$^2$High Energy Physics Department, Petersburg Nuclear
  Physics Institute, 188350 Gatchina, Russia}
\affiliation{$^3$Institut f\"ur Kernphysik, Forschungszentrum J\"ulich, 52425 J\"ulich, Germany}
\affiliation{$^4$High Energy Physics Institute, Tbilisi State University, 0186 Tbilisi, Georgia}
\affiliation{$^5$Institut f\"ur Kernphysik, Universit\"at zu K\"oln, 550937 K\"oln, Germany}
\affiliation{$^6$University College London, London WC1E 6BT, U.K.}

\date{\today}

\begin{abstract}
Indications for the production of a neutral excited hyperon in the
reaction $pp \rightarrow p K^+ Y^{0*}$ are observed in an experiment
performed with the ANKE spectrometer at COSY-J\"ulich at a beam
momentum of 3.65~GeV/c. Two final states were investigated
simultaneously, \emph{viz} $Y^{0*} \rightarrow \pi^+ X^-$ and $\pi^-
X^+$, and consistent results were obtained in spite of the quite
different experimental conditions. The parameters of the hyperon state
are $M(Y^{0*})= (1480\pm 15)$~MeV/c$^2$ and $\Gamma(Y^{0*})= (60\pm
15)$~MeV/c$^2$. The production cross section for $Y^{0*}$ decaying through these channels
is of the order of few
hundred nanobarns.  Since the isospin of the $Y^{0*}$ has not been
determined here, it could either be an observation of the
$\Sigma(1480)$, a one--star resonance of the PDG tables, or
alternatively a $\Lambda$ hyperon.  Relativistic quark models for the
baryon spectrum do not predict any excited hyperon in this mass range
and so the $Y^{0*}$ may be of exotic nature.
\end{abstract}
\maketitle

The question of how hadrons arise from QCD is central to a fundamental
understanding of hadronic multiquark and gluon systems. There has been
a recent renaissance of QCD spectroscopy, triggered by observations of
new narrow resonances, enhancements near thresholds, and possibly
exotic states. Taken in conjunction with lattice QCD, which is poised
to provide the theoretical insight into strong QCD, the new data may
pave the way to achieve this understanding.

The production of hyperons and their decay properties have been a
focus of experimental investigations ever since their discovery,
mostly in hadron-induced reactions, but recently also in
photoproduction. In comparison to the excitation spectrum of the
nucleon resonances ($N,\,\Delta$), the excited states of hyperons
($\Lambda,\, \Sigma$) are still much less well known. The nature of
experimentally well established states, such as the $\Lambda(1405)$,
is not at all understood yet. This hyperon may be a genuine three
quark system, a molecular-like meson-baryon bound state, or even of
exotic type.

The $\Sigma(1480)$ hyperon is far from being an established
resonance. In the most recent compilation of the PDG~\cite{PDG}, it is
described as a ``bump'', with unknown quantum numbers and given a mere
one--star rating. Very recently ZEUS has reported indications for a
structure in the invariant mass spectrum for $K^0_s p$ and $K^0_s
\bar{p}$, which may correspond to the
$\Sigma(1480)$~\cite{ZEUS_PLB591_7}. However, the structure appears on
a steeply varying background and therefore its significance is
difficult to estimate. The Crystal Ball investigation of the $K^- p
\rightarrow \pi^0 \pi^0 \Lambda$ reaction showed no sign for the
resonance in the $\pi^0 \Lambda$ invariant mass spectra, but it should
be noted that these are dominated by the
$\Sigma^0(1385)$~\cite{Crystal_PRC69_042202}.

In view of this uncertainty, we have investigated whether additional
information might be obtained from proton-proton interactions at low
energies. In so doing, we have found evidence for a neutral hyperon
resonance $Y^{0*}$ in data originally taken for scalar meson
production studies~\cite{a0_1, a0}.

The experiments have been performed at the Cooler Synchrotron COSY, a
medium energy accelerator and storage ring for protons and deuterons,
which is operated at the Research Center J\"ulich
(Germany)~\cite{COSY}. COSY supplied stored proton beams with a
momentum of 3.65~GeV/c at a revolution frequency of \mbox{$\sim10^{6} s^{-1}$}.
Using a hydrogen cluster--jet target (areal density
\mbox{$\sim 5 \times 10^{14}\,\textrm{cm}^{-2}$}) the average
luminosity during the measurements was $L=(1.38\pm 0.15)\times
10^{31}\,\textrm{s}^{-1}\,\textrm{cm}^{-2}$.

The ANKE spectrometer~\cite{ANKE_NIM} used in the experiments consists
of three dipole magnets, which guide the circulating COSY beam through
a chicane. The central C--shaped spectrometer dipole D2, placed
downstream of the target, separates the reaction products from the
beam. The ANKE detection system, comprising range telescopes,
scintillation counters and multi--wire proportional chambers,
simultaneously registers particles of either charge and measures their
momenta~\cite{K_NIM}.

A multibody final state, containing a proton, a positively charged
kaon, and a charged pion, together with an unidentified residue $X$,
was studied in the \mbox{$pp \rightarrow pK^+\pi^\pm X^\mp$}
reaction. Positively charged kaons and pions could be measured in the
momentum ranges \mbox{0.2--0.6~GeV/c} and 0.2--0.9~GeV/c, respectively,
negative pions between 0.4 and 1.0~GeV/c, and protons from 0.75~GeV/c
up to the kinematic limit. The angular acceptance of D2 is
$|\vartheta_{\mathrm H}| {\lesssim} 12^{\circ}$ horizontally and
~$|\vartheta_{\mathrm V}| {\lesssim} 5^{\circ}$ vertically for any
ejectile. By measuring delayed signals from the decay of stopped
kaons, $K^+$ mesons can be identified in a background of pions,
protons and scattered particles up to $10^6$ times more intense.

\begin{figure}[ht]
\vspace*{-0.5cm}
\psfig{file=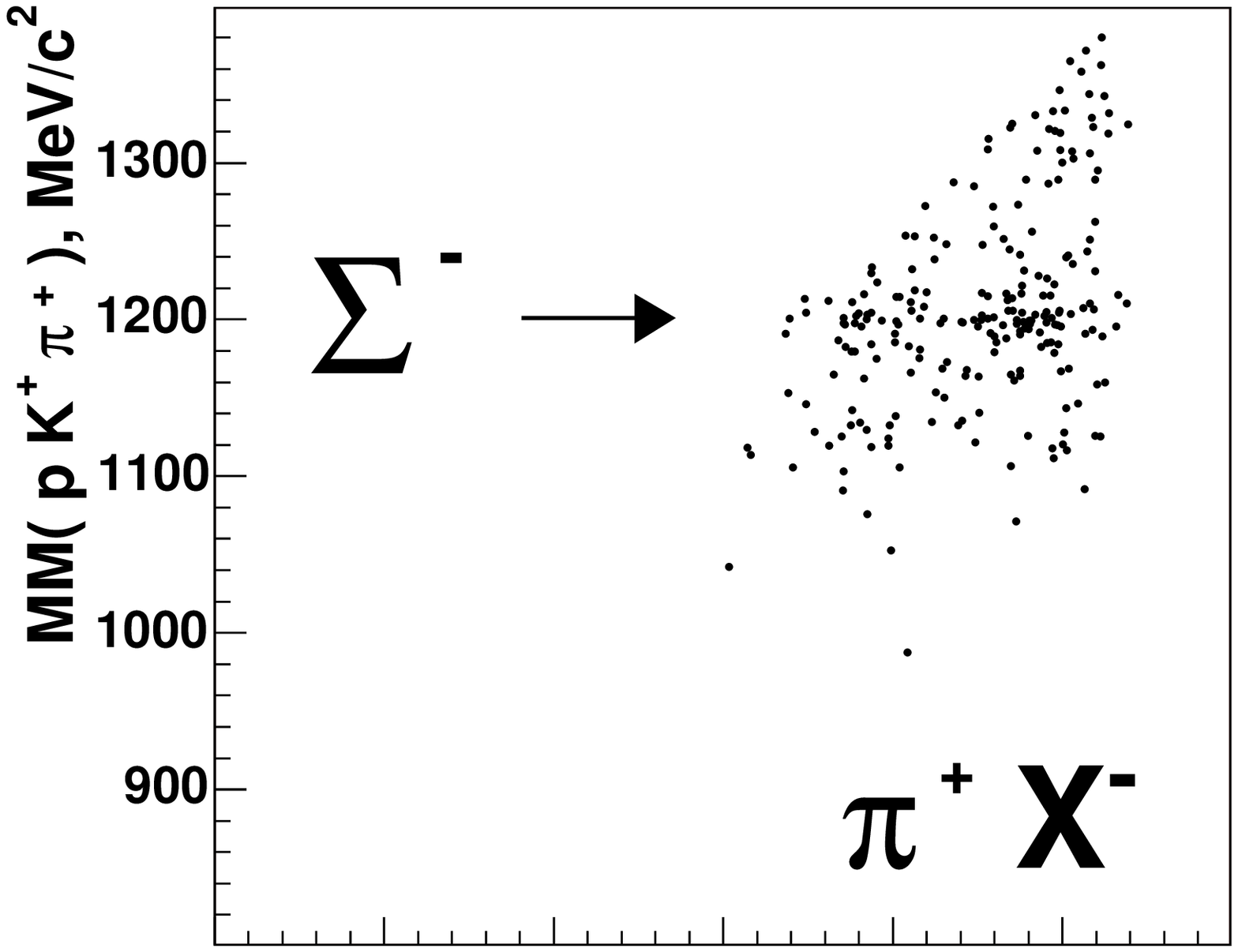,width=5.2cm}
\ \\
\hspace*{0.05mm}
\psfig{file=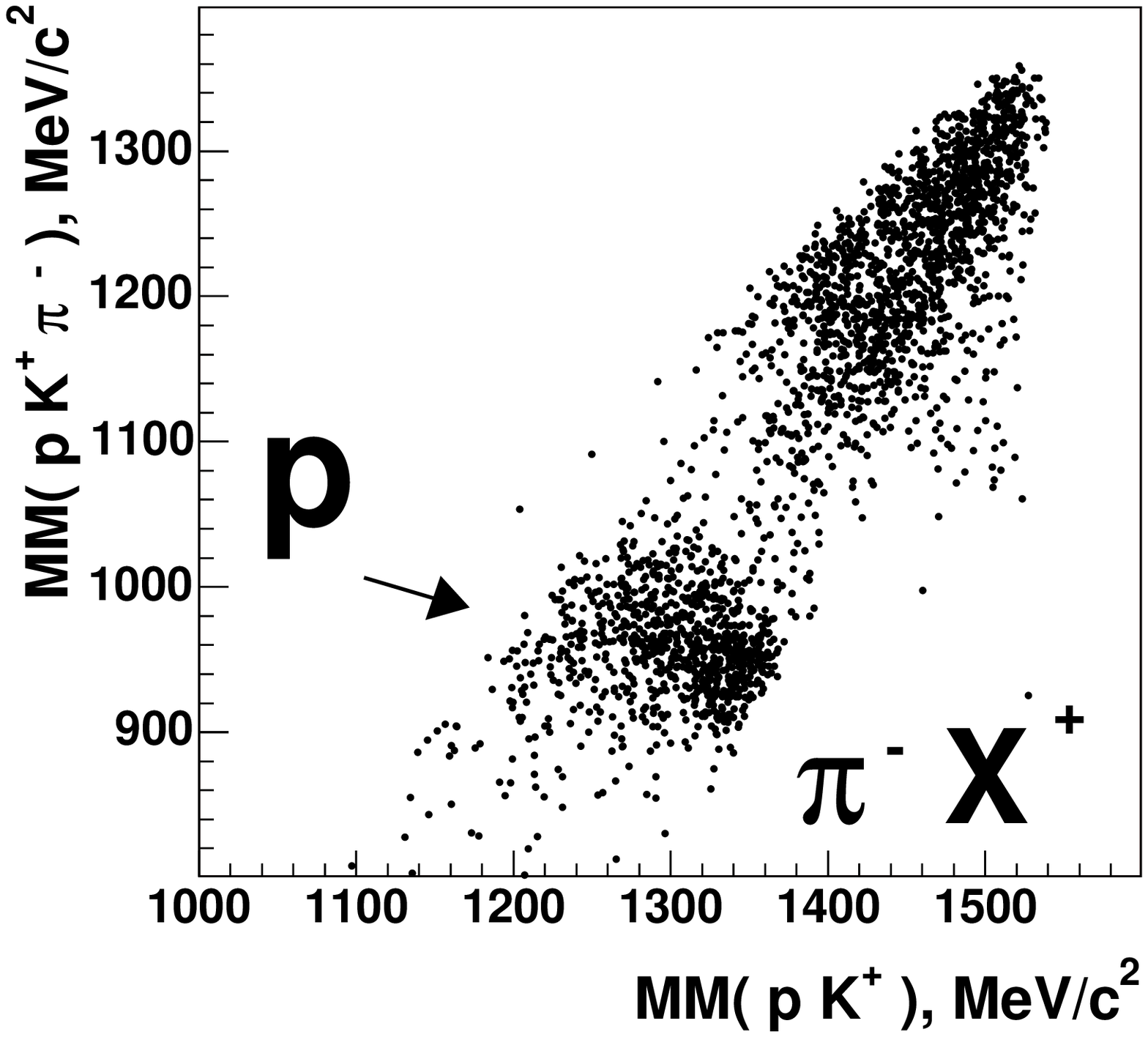,width=5.2cm}
 \vspace*{-0.4cm}
\caption{ Missing--mass $MM(pK^+\pi)$ vs $MM(pK^+)$
 distributions for $\pi^+$ (upper) and $\pi^-$ (lower) obtained in the
 reaction $pp \rightarrow pK^+\pi^\pm X^\mp$.
  }
  \label{fig:mm3_2}
\end{figure}

Events with three identified charged particles ($p,\,K^+,\,
\pi^{\pm}$) were retained for further analysis. In
Fig.\ref{fig:mm3_2} the missing--mass distributions $MM(pK^+\pi)$ vs
$MM(pK^+)$ are shown for the reaction channels $pp \rightarrow pK^+\pi^+X^-$
and $pp \rightarrow pK^+\pi^-X^+$. The triangular shape of the
distributions is due to the combination of kinematics and ANKE
acceptance. Since the probability for detecting three--particle
coincidences $(pK^+\pi^+)$ is about an order of magnitude smaller than
for $(pK^+\pi^-)$, the resulting numbers of events are also
drastically different.

\begin{figure}[ht]
\psfig{file=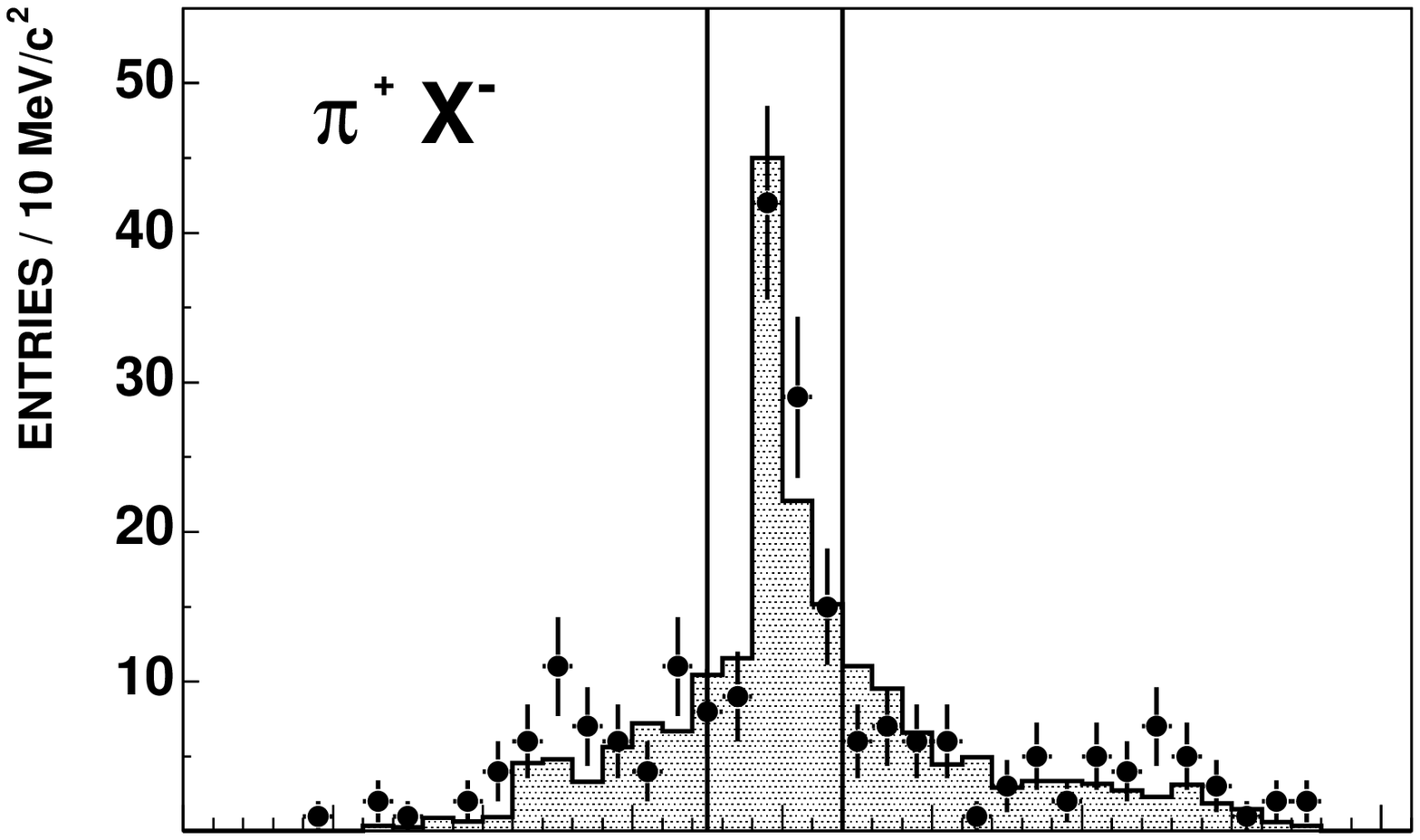,width=5.6cm}
\\
\vspace*{-1.33cm}
\psfig{file=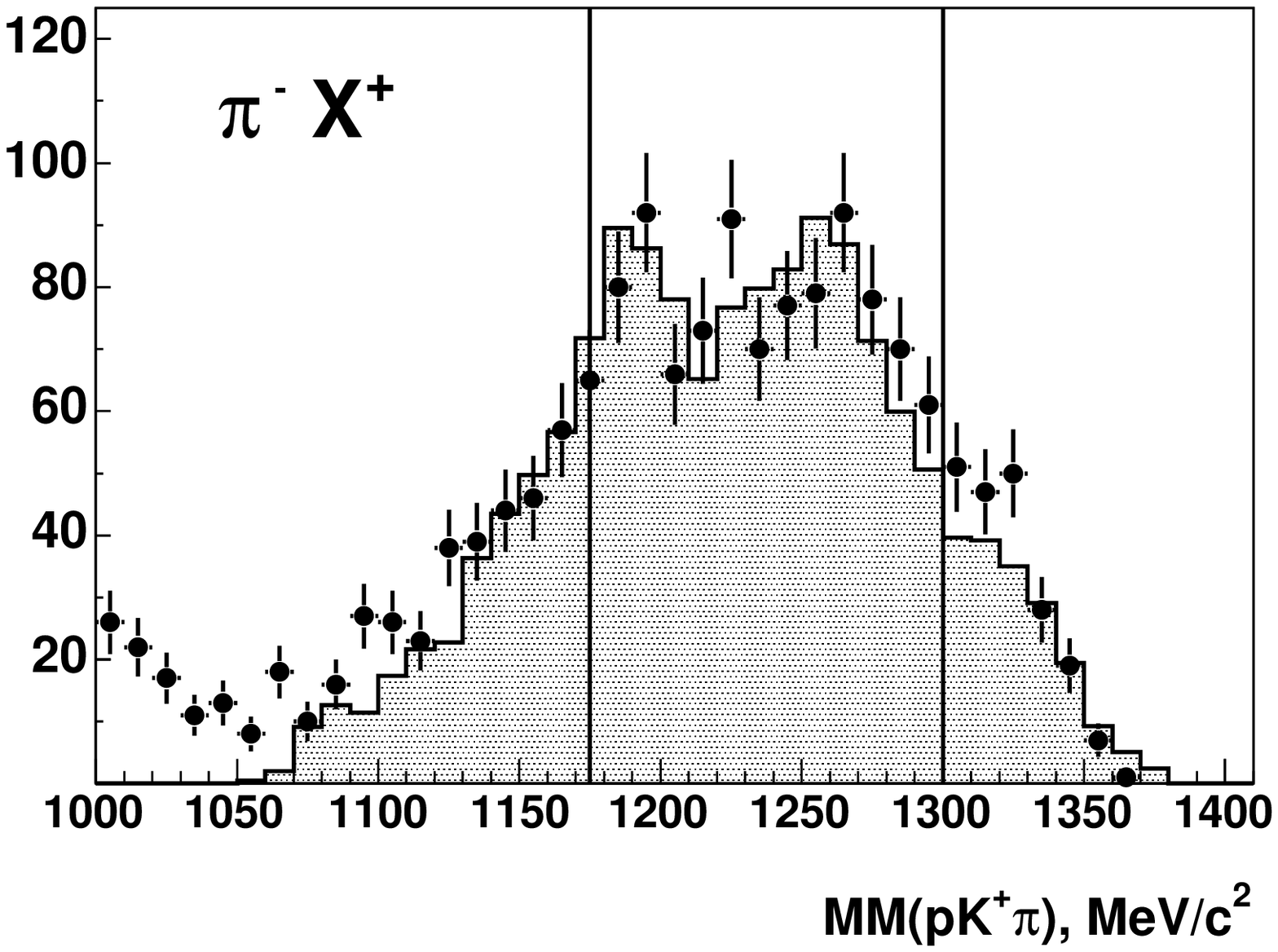,width=5.6cm}
  \caption{
Projections of Fig.\ref{fig:mm3_2} onto the three-particle missing
mass $MM(pK^+\pi^{\pm})$. Vertical lines show the $\Sigma$ bands
used for event selection. The results of simulations for
$MM(pK^+\pi^{\pm})>$1050~MeV/c$^2$, described in the text below,
are shown as a filled histogram.
  }
\label{fig:MM3}
\end{figure}

For the reaction $pp \rightarrow pK^+\pi^+X^-$ a clear
enhancement, corresponding to \mbox{$X^-=\Sigma^-(1197)$}, is
observed on a top of a low background (see projection of the upper
part of Fig.\ref{fig:mm3_2} in the upper part of
Fig.\ref{fig:MM3}). In the charge--mirrored \mbox{$pp \rightarrow
pK^+\pi^-X^+$} case, the $\pi^-$ may originate from different
sources, \textit{e.g.}\ a reaction with
\mbox{$X^+=\Sigma^+(1189)$} or a secondary decay of $\Lambda \to
p\pi^-$, arising from the major background reaction $pp
\rightarrow pK^+\Lambda \rightarrow pK^+\pi^-p$. Protons from the
latter reaction have been rejected by cutting $MM(pK^+\pi^-)$
around the proton mass (lower part in Fig.\ref{fig:mm3_2}).
Nevertheless the missing--mass distribution for the
$(\pi^-X^+)$-final state is more complicated and the
$\Sigma^+(1189)$ band is almost hidden underneath a strong
background of \textit{e.g.}\ $\pi^0 p$, $\pi^0\gamma p$, $\pi^+ n$
arising from the decay of heavier hyperons (see lower part of
Fig.\ref{fig:MM3}). For both final states, background due to
misidentified particles of different type is experimentally
estimated to be $<3\%$.

\begin{figure*}[t]
\psfig{file=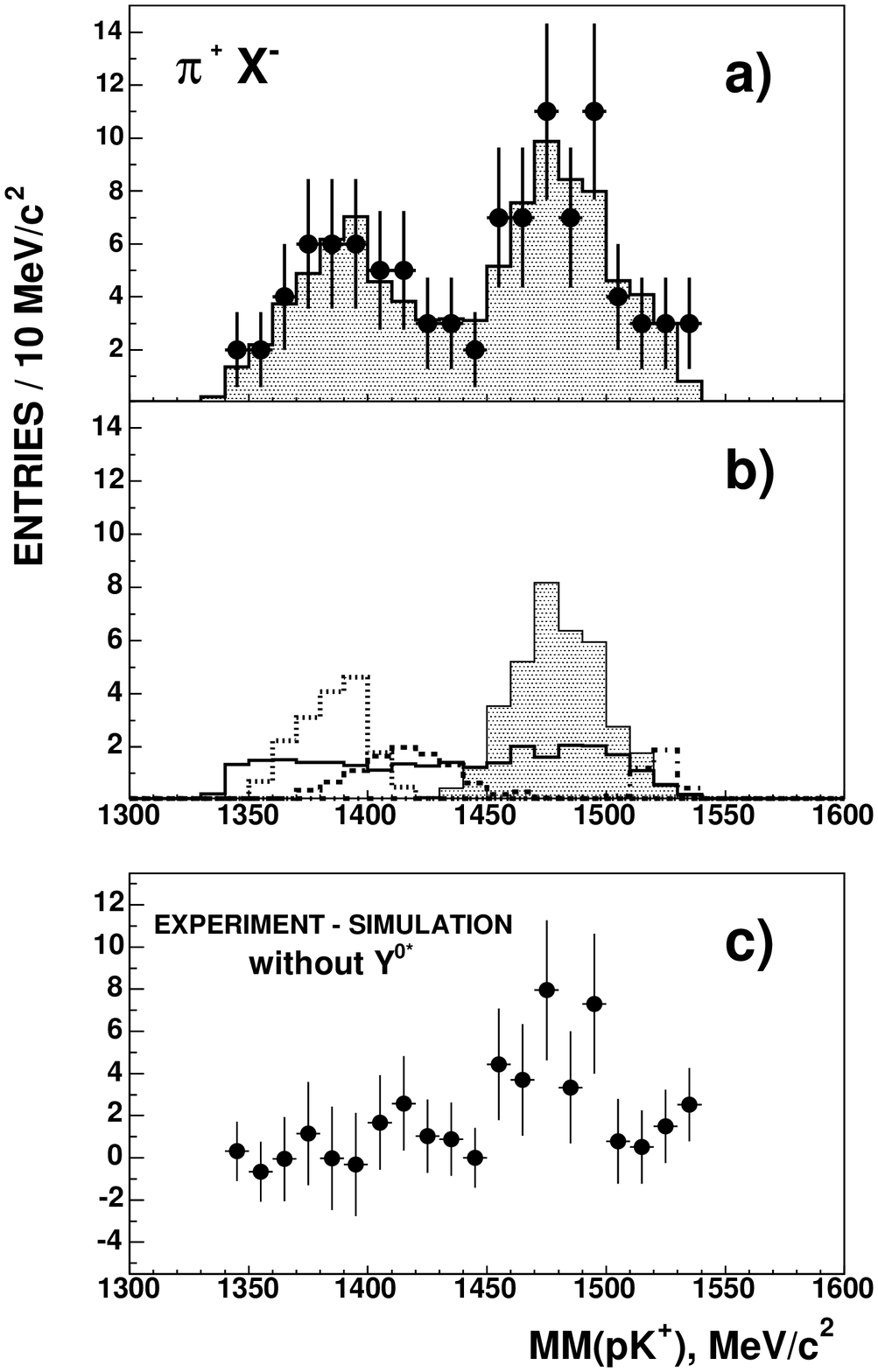,width= 8.6cm} \hspace*{-1.5cm}
\psfig{file=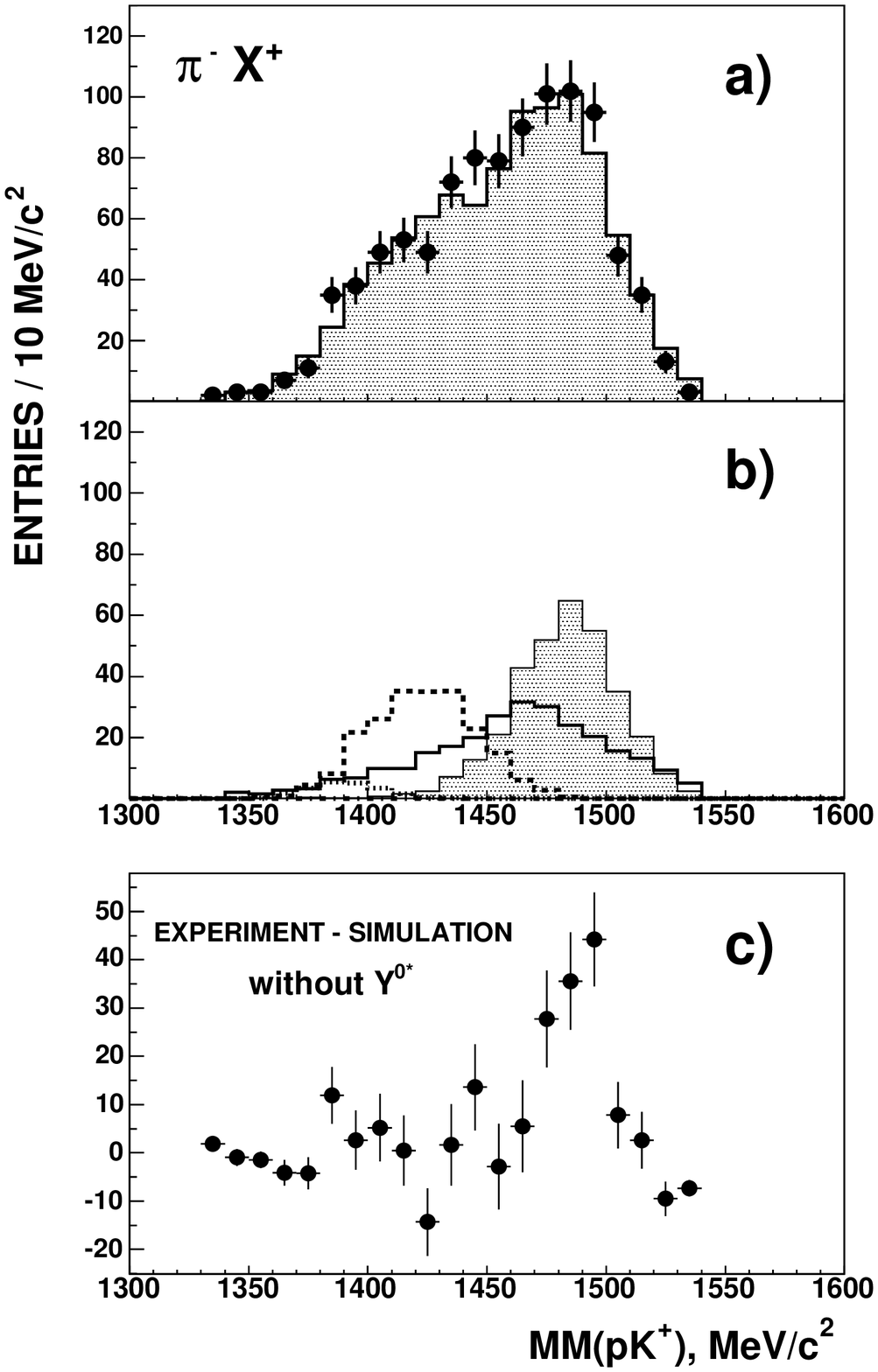,width= 8.6cm}
\hspace*{-0.0cm} \vspace*{-1.cm}
\caption{Missing--mass $MM(pK^+)$ spectra for the reaction $pp \rightarrow
    pK^+\pi^+X^-$(left) and $pp \rightarrow pK^+\pi^-X^+$(right). a)
    Experimental points with statistical errors are compared to the
    shaded histograms of the fitted overall Monte Carlo simulations;
    b) The simulation includes contributions from (i) resonances
    ($\Sigma(1385)$(dotted), $\Lambda(1405)$(dashed), $\Lambda
    (1520)$(dotted-dashed)), (ii) non-resonant phase-space production
    (solid), and (iii) the $Y^{0*}$ resonance (shaded histogram), as
    described in the text; c) Difference between the measured spectra
    and the sum of contributions (i)+(ii) fitted {\textit{without}}
    $Y^{0*}$ production. Note that the contributions of the individual
    partial channels are different for b) and c).}
\label{fig:final}
\end{figure*}

For further event selection, different cuts have been applied for
the two final states: for ($\pi^+X^-$) the $\Sigma^-$ has been
selected (1175 and 1220~MeV/c$^2$), while for ($\pi^-X^+$) the
corresponding range is between 1175 and 1300~MeV/c$^2$ in order to
include $\Sigma^+$ as well as $\Sigma^-$ with a $\pi^-$ in its
decay. This cut largely excludes neutral hyperons producing a
final state with two protons. The missing--mass distributions
$MM(pK^+)$ for such events are plotted in parts a) of
Fig.\ref{fig:final}. For $\pi^+X^-$, a double-humped structure is
observed, with peaks around $1390~\textrm{MeV/c}^2$ and
$1480~\textrm{MeV/c}^2$ (upper left).  In the $\pi^-X^+$ case, the
distribution also peaks at 1480~MeV/c$^2$ (upper right). The
different shapes and event numbers of the resulting spectra are
due to the various sources of $\pi^+$ and $\pi^-$ (see above). An
obvious question is whether these distributions can be explained
by the production of well established hyperon resonances
($\Sigma(1385), \Lambda(1405)$ and $\Lambda (1520)$) plus
non-resonant contributions or whether an additional source needs
to be invoked, \textit{e.g.}\ $pp \rightarrow pK^+Y^{0*}$ with a
further $Y^{0*}$ hyperon state.

In order to try to answer this question, Monte Carlo simulations
have been performed for both final states, using a simulation
package based on GEANT3~\cite{ANKE_GEANT}. The following reactions
with $(pK^+\pi^\pm)$ in the final state have been used as input
for this, assuming phase-space distributions and applying any
constraints due to isospin invariance:\\
(i) intermediate hyperon resonance production~\cite{PDG}
\vspace*{-0.3cm}
\begin{tabbing}
k1     \hspace*{2.5cm}\=       k2  \kill
$pp \rightarrow p{K}^+ \Sigma(1385)$ \> $\rightarrow pK^+ \pi^0 (\pi^- p)$\\
    \> $\rightarrow pK^+ \pi^\pm (\pi^\mp n)$\\
\>$\rightarrow pK^+ \pi^- (\pi^0 p)$\\
$pp \rightarrow p{K}^+ \Lambda(1405)$\> $\rightarrow pK^+ \pi^0 (\pi^- p) \gamma$\\
\>$\rightarrow pK^+ \pi^\pm (\pi^\mp n)$\\
\>$\rightarrow pK^+ \pi^- (\pi^0 p)$\\
$pp \rightarrow p{K}^+ \Lambda(1520)$\> $\rightarrow pK^+ p (\pi^+ \pi^- \pi^-)$\\
\>$\rightarrow pK^+ p (\pi^- \pi^0 \pi^0)$\\
\>$\rightarrow pK^+ n (\pi^+ \pi^-)$\\
\>$\rightarrow pK^+ \pi^0 (\pi^- p) \gamma$\\
\>$\rightarrow pK^+ \pi^+ \pi^- (\pi^- p)$\\
\>$\rightarrow pK^+ \pi^+ \pi^- (\pi^0 n)$\\
\>$\rightarrow pK^+ \pi^\pm (\pi^\mp n)$\\
\>$\rightarrow pK^+ \pi^- (\pi^0 p)$
\end{tabbing}
\vspace*{-0.25cm}
(ii) non-resonant production\\
$pp \rightarrow NK^+X$\\
$pp \rightarrow NK^+\pi X$\\
$pp \rightarrow NK^+\pi\pi X$,\\
with constraints for the relative contributions obtained on the
basis of measured cross sections and phase space considerations in
the non-resonant cases~\cite{Baldini, Byckling}.

The final state of ($\pi N$) results from the decay of ground-state hyperons, while
($2\pi$) and ($3\pi$) are from $K^0$ and $K^-$ decays,
respectively. $X$ represents any known $\Lambda$ or $\Sigma$ hyperon
which could be produced in the experiment. When there are two
particles of the same kind, both are further processed as in the analysis
of experimental data.

In the lower parts of Fig.\ref{fig:final}, the difference between the
measured missing--mass distributions and the sum of fitted resonant and
non-resonant production --- contributions listed under (i) and (ii) ---
is shown. For both final states ($\pi^+X^-$ and $\pi^-X^+$) the shape
of the measured distributions cannot be reproduced by the simulations
and an excess of events is observed around the missing mass of
$1480~\textrm{MeV/c}^2$ in both cases. It is therefore suggested that another
excited hyperon is produced and observed through the decay $pp
\rightarrow p{K}^+Y^{0*} \rightarrow pK^+ \pi^\pm X^\mp$.

The $Y^{0*}$ mass and width have been determined from a fit based on
simulations that cover the range from 1460 to 1490~MeV/c$^2$ and from
45 to 75~MeV/c$^2$, respectively, both in steps of 5~MeV/c$^2$. From a
minimization procedure the following parameters of the $Y^{0*}$,
consistent for both final states, are obtained: $M(Y^{0*})=
(1480\pm15)~\textrm{MeV/c}^2$ and $\Gamma(Y^{0*}) = (60\pm
15)~\textrm{MeV/c}^2$. Note that the experimental mass resolution is
of the order of 10~MeV/c$^2$.

The fits to the data are shown in the two parts of
Fig.\ref{fig:final}a), while the individual contributions are plotted
in parts b).
These contributions are also used to obtain the three-particle
missing--mass spectra for $MM(pK^+\pi^{\pm})>$1050~MeV/c$^2$ as shown in Fig.\ref{fig:MM3}:
in comparison with the experimental results a good agreement is achieved.

The numbers of events in the two peaks are:
$S(Y^{0*}\rightarrow\pi^+X^-)=35\pm10$ and
$S(Y^{0*}\rightarrow\pi^-X^+)=330\pm60$. The statistical significance
of the signal, assuming that this is due to the production of the
$Y^{0*}$, is
at least 4.5~standard deviations.

In order to estimate the production cross section for $Y^{0*}$ decaying through these channels,
we used an overall detection efficiency of $\sim$7\% and an integrated luminosity
of $\sim$6~pb$^{-1}$.
With acceptances obtained from simulations presented above,
we arrive at cross section values of (450$\pm$150$\pm$150)~nb
for ($\pi^+X^-$) and (1200$\pm$250$\pm$500)~nb for ($\pi^-X^+$).
The first error is statistical, while the second represents the systematic
uncertainty.
It can thus be concluded that the cross section estimates
are consistent for both final states and are of the order of few
hundred nanobarns.

Assuming that the $Y^{0*}(1480)$ hyperon exists, we briefly address
the question of its possible theoretical description. In the
constituent quark model, baryons are interpreted as bound states of
three valence quarks where hyperons contain at least one strange
quark. The baryon spectrum has been investigated systematically in a
relativistic quark model with instanton-induced quark forces. No
excited $\Lambda$ or $\Sigma$ resonances, in addition to the well
known states, have been found for masses below $\approx
1600~\textrm{MeV/c}^2$~\cite{EPJA10_447}. These findings are in
agreement with results obtained in the relativized quark model of
Capstick and Isgur~\cite{CapIs}. Thus, it seems to be difficult to
reconcile the low mass of the $Y^{0*}$ within the existing
classification of 3q-baryons.

In early papers, configurations of four quarks and an antiquark
have been discussed for this mass range. H\"ogassen and Sorba have performed a
group-theoretical classification of such states and arrived at an
estimate of 1440~MeV/c$^2$ for the mass of an exotic
$\Sigma$~\cite{NPB145_119}. Azimov et al.  introduced a flavor octet
and an antidecuplet of exotic baryons~\cite{Azimov_1970, Azimov}.  For
the octet, a $\Sigma$ state with a mass of 1480~MeV/c$^2$ and a
$\Lambda$ state at 1330~MeV/c$^2$ have been suggested in
Ref.~\cite{Azimov}. There were also attempts to mix octet and
antidecuplet states, based on diquark correlations, as proposed by
Jaffe and Wilczek~\cite{Jaffe_Wilczek_PRD69_114017}. A $\Sigma$
resonance with a mass of 1495~MeV/c$^2$ has been predicted as a member
of the mixed multiplet~\cite{PRD69_094009_antidecuplet}.

In models, which couple nucleons with kaons and pions, quasibound
states are generated with relatively low masses. In
Ref.~\cite{PLB582_49_quark_mass}, a pole with the quantum numbers of
the $\Sigma$, which might be identified with the $\Sigma(1480)$, is
found at a mass of 1446~MeV/c$^2$ though a width of 343~MeV/c$^2$ is
much larger.

Since a clear theoretical picture has not yet appeared, any conclusion
about the nature of the $Y^{0*}$ would be premature.

In summary, we have observed indications in proton--proton collisions
at 3.65~GeV/c for a neutral hyperon resonance $Y^{0*}$ decaying into
$\pi^+ X^-$ and $\pi^- X^+$ final states. Its parameters are
$M(Y^{0*}) = (1480\pm 15)~\textrm{MeV/c}^2$ and $\Gamma(Y^{0*}) =
(60\pm 15)~\textrm{MeV/c}^2$ though, since it is neutral, it can be
either a $\Lambda$ or $\Sigma$~hyperon. The production cross section
is of the order of few hundred nanobarns.  On the basis of existing
data we cannot decide whether it is a three--quark baryon or an exotic
state, although some preference towards its exotic nature may be
deduced from theoretical considerations.

Further studies are required to confirm the existence of the
$Y^{0*}(1480)$ hyperon and to determine its quantum numbers.  Such
measurements, in particular for $Y^*$ decays with photons in the final
state are foreseen with the WASA detector at
COSY~\cite{WASA}. Searches for the charged $Y^{-*}$ hyperon in the
reaction $pn\rightarrow pK^+ Y^{-*} \rightarrow pK^+
\pi^- X^0$, using a deuterium cluster-jet target and spectator proton
tagging, are also conceivable.

We acknowledge contributions by P.~Kravtchenko for early simulations,
to D.~Gotta and B.~Nefkens for critical discussions. We
also thank all other members of the ANKE collaboration and the COSY
accelerator staff for their help during the data taking.

This work has been supported by: FFE Grant (COSY-78, nr 41553602),
BMBF (WTZ-RUS-211-00, 691-01, WTZ-POL-015-01, 041-01), DFG (436 RUS
113/337, 444, 561, 768), Russian Academy of Sciences (02-04-034,
02-04034, 02-18179a, 02-06518).

{$^*$ present address: Physikalisches Institut, Universit\"at Bonn, Nuss\-allee 12, 53115 Bonn, Germany}

\end{document}